\documentclass[preprint2]{aastex}

	\newcommand{\co}{$^{12}\mathrm{CO}$ }

\shorttitle{Gas inside the 97 au cavity around the transition disk Sz\,91}
\shortauthors{Canovas et al.}

\begin{document}


\title{Gas inside the 97 au cavity around the transition disk Sz\,91}


\author{H. Canovas\altaffilmark{1,11}, 
M.~R. Schreiber\altaffilmark{1,11}, 
C. C\'aceres\altaffilmark{1,11},
F. M\'enard\altaffilmark{2}, 
C. Pinte\altaffilmark{2},
G.~S. Mathews\altaffilmark{3,4}, 
L. Cieza\altaffilmark{5,11},
S. Casassus,\altaffilmark{6,11},
A. Hales\altaffilmark{7,8},
J.~P. Williams\altaffilmark{9},
P. Rom\'an\altaffilmark{10,11},
A. Hardy\altaffilmark{1,11}
}
\affil{Departamento de F\'isica y Astronom\'ia, Universidad de Valpara\'iso, Valpara\' iso, Chile}
\email{hector.canovas@dfa.uv.cl}

%
%


\altaffiltext{2}{UMI-FCA, CNRS/INSU, France (UMI 3386), and Dept. de Astronom\'{\i}a, Universidad de Chile, Santiago, Chile.}
\altaffiltext{3}{Leiden Observatory, Leiden University, PO Box 9513, 2300 RA Leiden, the Netherlands}
\altaffiltext{4}{Department of Physics and Astronomy, University of Hawaii, Honolulu, HI 96822 USA}
\altaffiltext{5}{Facultad de Ingenier\'ia, Universidad Diego Portales, Av. Ejercito 441, Santiago, Chile}
\altaffiltext{6}{Departamento de Astronom\'ia, Universidad de Chile, Casilla 36-D, Santiago, Chile}
\altaffiltext{7}{Atacama Large Millimeter/Submillimeter Array, Joint ALMA Observatory, Alonso de C\'ordova 3107, Vitacura 763-0355, Santiago - Chile}
\altaffiltext{8}{The National Radio Astronomy Observatory is a facility of the National Science Foundation operated under cooperative agreement by Associated Universities, Inc}
\altaffiltext{9}{Harvard-Smithsonian Center for Astrophysics, 60 Garden Street, Cambridge, MA 02138, USA}
\altaffiltext{10}{Center of Mathematical Modeling, Universidad de Chile, Beauchef 851, Santiago, Chile.}
\altaffiltext{11}{Millennium Nucleus ``Protoplanetary Disks in ALMA Early Science"}


\begin{abstract}
We present ALMA (Cycle 0) band-6 and band-3 observations of the transition disk Sz\,91. The disk inclination and position angle
are determined to be $i=49.5\degr\pm3.5\degr$ and $\mathrm{PA}=18.2\degr\pm3.5\degr$ and the dusty and gaseous disk are
detected up to $\sim220$ au and $\sim400$ au from the star, respectively. Most importantly, our continuum observations
indicate that the cavity size in the mm-sized dust distribution must be $\sim97$ au in radius, the largest cavity observed around a T
Tauri star. Our data clearly confirms the presence of \co(2-1) well inside the dust cavity. Based on these observational constrains
we developed a disk model that simultaneously accounts for the \co and continuum observations (i.e., gaseous and dusty disk).
According to our model, most of the millimeter emission comes from a ring located between 97 and 140 au. We also find
that the dust cavity is divided into an innermost region largely depleted of dust particles ranging from the dust sublimation radius
up to 85 au, and a second, moderately dust-depleted region, extending from 85 to 97 au. The extremely large size of the dust
cavity, the presence of gas and small dust particles within the cavity and the accretion rate of Sz\,91 are consistent with the
formation of multiple (giant) planets. 
\end{abstract}




\section{Introduction}

Simulations of disk-planet interactions show that a giant planet embedded in a circumstellar disk should open a gap
or carve an inner hole in the disk \citep{lin1986}. These gaps or holes are imprinted in the spectral energy distributions
(SEDs) of young stellar objects as reduced near and/or mid infrared excess emission. Protoplanetary disks showing
this observational feature are usually called ``transition disks", although different names and classifications are found
in the literature \citep{evans2009}. Apart from planet formation, three alternative mechanisms can create opacity holes
in the inner disk: grain growth \citep{dullemond2005}, photoevaporation \citep{alexander2006}, and tidal truncation by
close stellar companions \citep{artymowicz1994}. 

The latest models of photoevaporation show that only transition disks with small accretion rates and/or small
inner cavities can be explained by this process \citep{owen2012}. While grain growth and transport effects can
explain the dips in the SEDs of transition disks, the most recent models fail in reproducing the large cavities observed
in millimeter images of accreting transition disks \citep{birnstiel2013a}. Current planet formation models predict that a
forming planet may reduce (or remove if the planet is massive enough) the amount of small particles within the disk's
cavity while allowing the gas to flow through the cavity \citep[e.g., ][]{rice2006, lubow2006,dodson2011, zhu2011}.

The transition disk Sz\,91, at 200 pc \citep{merin2008}, has been classified as a planet forming candidate based on its
lack of detected companions down to 30 au, its disk mass, accretion rate ($7.9\times10^{-11}\mathrm{M_{\sun}}\,\mathrm{yr}^{-1}$),
and SED shape \citep{romero2012}. Recently,
\citet{tsukagoshi2014} (hereafter T2014) resolved the inner cavity in K-band polarized light with Hi CIAO/Subaru,
and find evidences for a gap in the continuum at 345 GHz (870 \micron) from SMA observations. These authors also
detect the outer gaseous disk using the $^{12}\mathrm{CO}(3-2)$ line. We here use ALMA cycle 0 band-6 and band-3
observations to describe the disk in more detail. Our observations reveal a huge cavity of $\sim$97 au in radius
in the continuum at 230 GHz (1.3 mm), and confirm the presence of $^{12}\mathrm{CO} (2-1)$ inside the cavity. We also
derive stronger constrains on the disk's geometry and orientation.
\section{Observations and Data reduction}
\begin{table*}[t]
\begin{center}
\caption{Observing log.}
\label{tab:tab1}
\begin{tabular}{c c c c c c c c c}
\tableline\tableline
LO                 &      Date                     &     Flux                &     $\tau_{\nu}$  &     PWV      &      $\mathrm{t_{exp}}$ &  $\mathrm{T_{sys}}$	& bandwidth    \\ 
$\mathrm{[GHz]}$&  [dd/mm/yy]        &     Calibrator       &     Zenith            &     [mm]      &           [min]                &   [K]                 &  [GHz]                        \\	
\tableline
231.46           &     18/06/2012            &     Titan               &     0.09	           &     1.586      &           3.77                &     96                & $2\times$1.875         \\ 
231.46           &     02/07/2012            &     Neptune         &     0.05	           &     0.827      &           3.77                &     70                & $2\times$1.875         \\  
231.46           &     03/07/2012            &     Neptune         &     0.08               &     1.520      &           3.77                &     79                & $2\times$1.875         \\ 
231.46           &     04/07/2012            &     Titan               &     0.10               &     1.877      &           3.77                &     82                &$2\times$1.875         \\ 
109.99           &     01/08/2012            &     Neptune         &     0.03               &     1.552      &           0.4                  &     53                &$4\times$1.875           \\ 
\tableline
\end{tabular}
\end{center}
\end{table*}
\subsection{Interferometry}
Sz\,91 was observed with ALMA band-6 (231 GHz) and band-3 (110 GHz) during Cycle-0 (program 2011.0.00733,
PI: M.R. Schreiber). The observations are summarized in  Table~\ref{tab:tab1}.

Sz\,91 was observed in five different epochs with different weather conditions and antenna configurations for the band-6
data. The system temperature ranged from 53 to 96 K.  For the band-3 data, the ALMA correlator was configured to provide
two spectral windows centered on the continuum, and two spectral windows centered on the $^{13}\rm{CO(1-0)}$ (110.20135 GHz)
and the $\rm{C^{18}O(1-0)}$ (109.78218 GHz) lines. The total bandwidth of the band-3 observations was 7.5 GHz
($4\times1.875$ GHz). For the band-6 observations, the ALMA correlator was initially configured to provide 1 spectral window
centered on the continuum, and three spectral windows centered on the $^{12}\mathrm{CO}(2-1)$, 230.538 GHz (hereafter \co),
$^{13}\mathrm{CO}(2-1)$ and $\mathrm{C^{18}O}(2-1)$ lines. Unfortunately, because of technical problems only one sideband
of the correlator could be configured, and only the spectral windows centered on the continuum and on the \co line could be
produced. The total bandwidth of the band-6 observations was $2\times1.875$ GHz. Both ALMA bands were sampled at 0.488 MHz
(e.g., 0.635 km/s in the \co line).

The quasar QSO B1730-130 and the AGN ICRF J160431.0-444131 were used as bandpass and primary phase calibrator in
both bands, respectively. Neptune and Titan were used to calibrate in amplitude in band-3 and band-6, respectively. The
observations of the calibrators were alternated with the science observations. The individual exposure times for the calibrators
and the science targets was 6.05 seconds, amounting to a total exposure time for the science observations of 226 seconds
(3.77 minutes) in band-6 and 24 seconds (0.4 minutes) in band-3. Phase correction based on WVR measurements was
performed in offline mode, as part of the basic ALMA Cycle 0 corrections. We used the dispersion of the band-6 flux
calibrations to estimate a flux uncertainty of $\approx15\%$. This value is a lower limit since it does not include uncertainties
from the amplitude calibration\footnote{According to the documentation found in http://www.almaobservatory.org/ the amplitude
calibration uncertainties are $\lesssim5\%$.}.

The observations were processed using the Common Astronomy Software Application (CASA) package \citep{McMullin2007}.
The visibilities were Fourier transformed and deconvolved, using natural weighting, with the CLEAN algorithm \citep{Hogbom1974}.
For the band-6 data, we combined the four datasets to increase the $uv$ coverage before cleaning. The cleaned image produced
from this concatenated dataset had lower rms than the images created from individually cleaned datasets. After combining the
band-6 data, the two bands contained 276 baselines, ranging from 21 to 402 m (16 to 309 k$\lambda$) in band-6, and from 21
to 452 m  (7 to 167 k$\lambda$) in band-3.

In the continuum, we reach a rms of 0.1 mJy/beam in band-6 and 0.1 mJy/beam in band-3. The median rms on the channels
associated to the \co, is 6.2 mJy/beam. In the continuum, the beam is $0\farcs86 \times 0\farcs60$ with a position angle (PA)
of $88.9\degr$ in band-6, and  $2\farcs08 \times 1\farcs51$ with a PA of $99.4\degr$ in band-3. No emission was detected in
either the $^{13}\mathrm{CO}(1-0)$ or the $\mathrm{C^{18}O}(1-0)$ lines in band-3.

\subsection{Photometry}
We add two WISE flux at 12 and 22 \micron, two Herschel/PACS fluxes at 100 and $160\micron$, three Herschel/Spire
fluxes at 250, 350 and $500\micron$ and the two mm-fluxes discussed here to the photometry presented by \citet{romero2012}.
For the Herschel fluxes we adopt the most recent values presented by \citet{bustamante2015}. The fluxes at wavelengths
shorter than 24 $\micron$ were corrected of extinction by applying the dereddening relations listed in \citet{cieza2007}. We assume
calibration errors of 20\% for the optical photometry and 15\% for the photometry up to 24\micron. The photometry
between 100-500 $\micron$ is affected by background contamination from the cloud in which Sz\,91 is embedded, specially
at 250, 350 and 500 $\micron$ \citep{bustamante2015}. We therefore consider the Herschel fluxes at 250, 350 and 500 as
upper limits. For the 870$\micron$ flux we used the $8.5\%$ uncertainty listed by \citet{romero2012}. To derive the fluxes from
our ALMA observations we perform a fit in the $uv$-plane. Because of the lack of data at short baselines (i.e., large spatial scales),
deriving the total flux from the cleaned images would result in underestimated values. Instead we fit different profiles to the visibilities
to estimate the flux at the center of the $uv$-plane. In band-6 Sz\,91 is resolved (see next section), so we fit a Gaussian profile;
in band-3 Sz\,91 is not resolved, so we use a point-source profile. The fluxes comprising the SED, their errors and references,
are listed in Table~\ref{tab:tab2}. 
\begin{table}[t]
\begin{center}
\caption{Photometry of Sz\,91. All fluxes are extinction-corrected. References: (1)  \citet{romero2012},
(2) \citet{wright2010}. (3) \citet{merin2008}, (4) \citet{bustamante2015}}
\label{tab:tab2}
\begin{tabular}{c c c c}
\tableline\tableline
Wavelength		&	Flux				&	Error	&	Reference	\\ 
$[\mu$m]		&	[mJy]			     &	[mJy]	&				\\ 
\tableline
   0.65    &        37.1    &         7.4    &            1\\
   1.25    &        97.7    &        14.7    &           1 \\
   1.60    &       120.6    &        18.1    &          1 \\
   2.20    &        90.6    &        13.6    &           1 \\
   3.60    &        41.6    &         6.2    &            1 \\
   4.50    &        26.0    &         3.9    &            1 \\
   5.60    &        17.8    &         2.7    &            1 \\
   8.00    &        11.1    &         1.7    &            1 \\
  12.00    &         6.9    &         1.0    &            2 \\
  22.00    &         9.0    &         1.4    &            2 \\
  24.00    &         9.7    &         1.5    &            3 \\
  70.00    &       510.0    &       130.0    &        4 \\
 100.00    &       680.0    &       170.0    &       4 \\
 160.00    &       720.0    &       180.0    &       4 \\
 250.00\tablenotemark{a}    &       860.0    &       170.0    &       4 \\
 350.00\tablenotemark{a}    &       620.0    &       120.0    &       4 \\
 500.00\tablenotemark{a}    &       380.0    &        80.0    &        4 \\
 850.00    &        34.5    &         2.9    &          1 \\
1300.00    &        12.7    &         1.9    &          This work \\
2700.00    &         0.7    &         0.1    &           This work \\
\tableline
\end{tabular}
\tablenotetext{a}{Considered as an upper limit due to cloud emission.}
\end{center}
\end{table}

\section{Continuum and $\mathrm{^{12}CO}$ images}

In band-3, Sz\,91 is only detected (and not resolved) in the continuum, with a maximum signal to noise ratio (S/N) of $\approx4.6$
and a disk emission of $0.7\pm0.1$ mJy. The median rms in the individual channels is 11.0 mJy/beam. The band-6 results are
detailed below.

\begin{figure*}[t]
\center
\includegraphics[width = 1.0\linewidth,trim = 100 20 100 150]{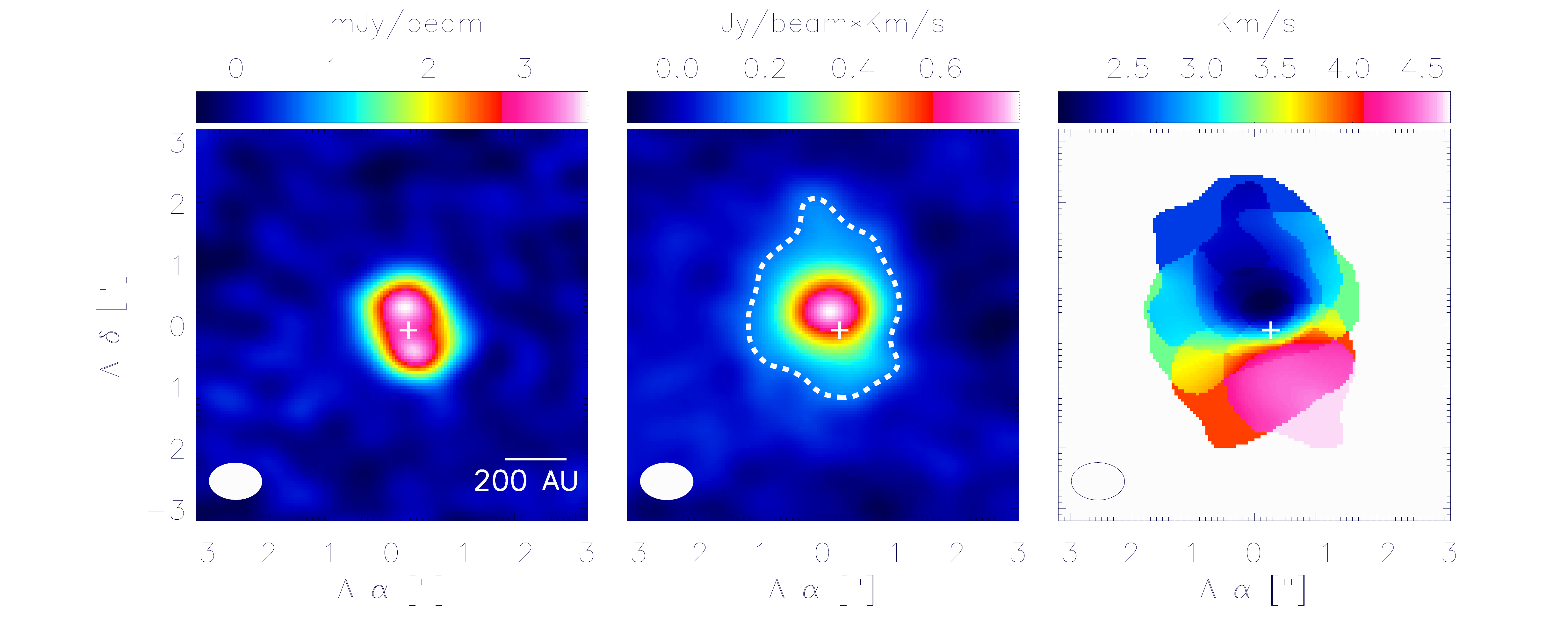}
\caption{Left: band-6 continuum image. Center: Moment-0 image, constructed from the \co line channels showing emission above the
$3\sigma$ level. The white dashed line contours the regions of the image with S/N larger than 3. The strong north-south asymmetry is
most likely caused by cloud contamination (see text). Right: The moment-1 image, constructed using the same channel images as for
the moment-0 image. A rotating gaseous disk is clearly identified. In all images, north is up, east is left. The white cross shows the disk's
center as derived in Sect. 4.3.}
\label{fig:fig1}
\end{figure*}

\begin{figure*}[t]
\center
\includegraphics[width = 1.0\linewidth,trim = 0 60 40 150]{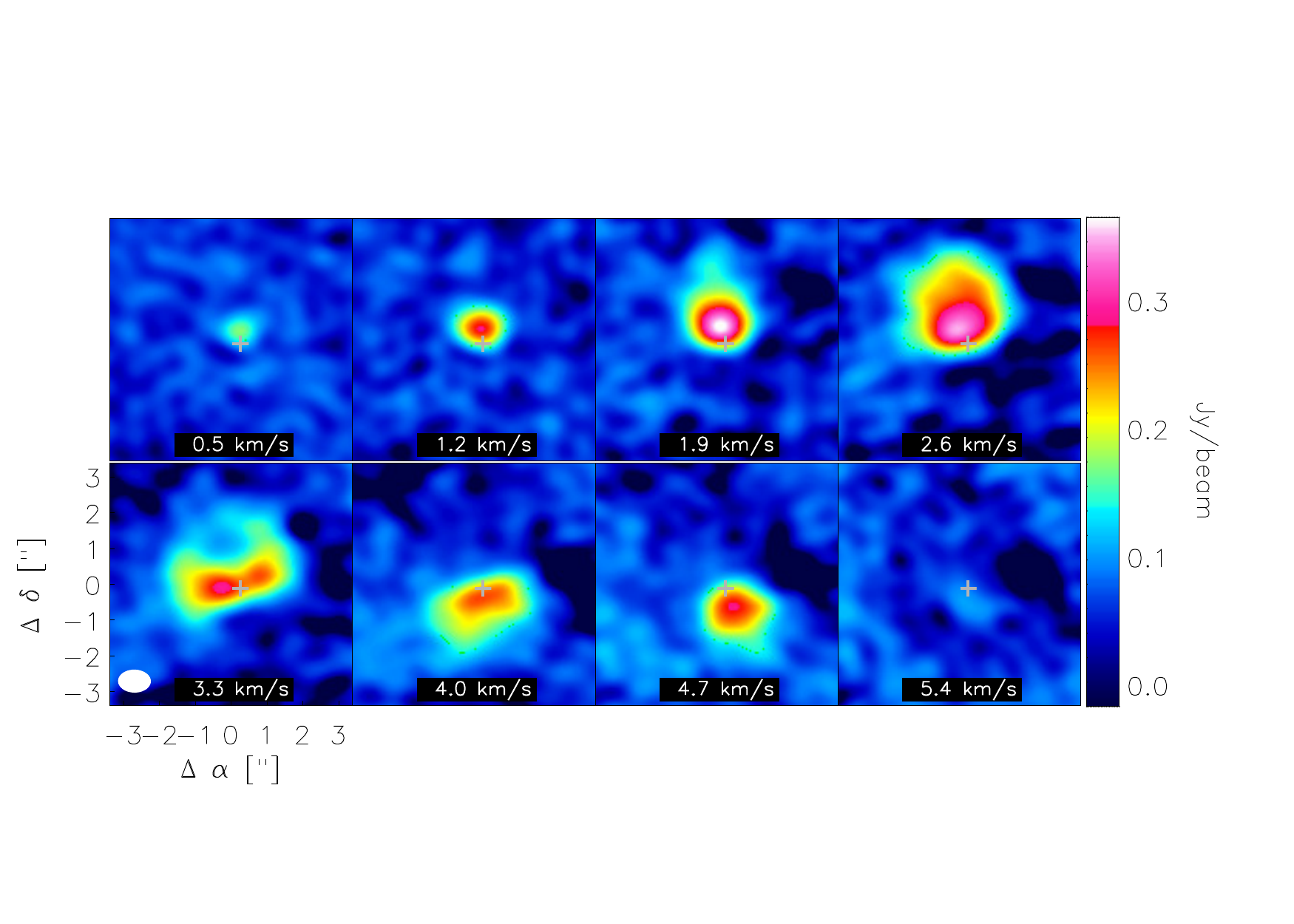}
\caption{Band-6 channels showing significant (above $3\sigma$) \co emission. The $v_\mathrm{LSR}$ velocity of each channel is
indicated in the legends. The channel at $v_\mathrm{LSR}=5.4$ km/s does not shows \co emission, and is shown for completeness.
The images are plotted in square root strecht. North is up an east is left. The channel at $v_\mathrm{LSR}=3.3$ km/s is the closest
to the derived systemic velocity of the system (see text). The grey cross shows the disk's center as derived in Sect. 4.3.}
\label{fig:fig2}
\end{figure*}

The continuum image (Fig.~\ref{fig:fig1}, left panel) shows an inclined disk (the geometry is derived in Sect. 3.3) with evidences
of an inner cavity, as expected from the disk's SED and the previous results of T2014. The inner hole is resolved along the
projected major axis of the disk. Integrating over the regions of the image with S/N higher than 3 results in $\mathrm{F_{1.3mm}} = 10.7\pm1.6$
mJy, which is $\sim 20\%$ lower than the integrated flux estimated from the visibilities (see Table~\ref{tab:tab2}). The peak flux
in the cleaned image is $3.7\pm0.5$ mJy/beam. The difference in brightness between the peak flux of the northern and
southern lobes is below 3$\sigma$.
The channels corresponding to $v_\mathrm{LSR}$ velocities ranging from 
0.5 to 4.7 km/s show significant (above $3\sigma$) \co emission (Fig.~\ref{fig:fig2}). 

Note that the emission at 0.5 km/s, although faint when compared to the maximum emission of the \co line, is detected with
S/N $\approx6$. The corresponding moment 0 and moment 1 images are shown in Fig.~\ref{fig:fig1} (middle and right panel).
The moment 0 image shows a strong north-south asymmetry, which is most likely caused by emission from the cloud in which
Sz\,91 is embedded. Indeed, the systemic velocity of the cloud has been measured to be $v_\mathrm{LSR}\approx4.8$ km/s
\citep{vilasboas2000, tachinara2001}, which coincides with the redshifted $\mathrm{^{12}CO}$ emission of Sz\,91 (southern
lobe, see the moment-1 image in Fig.~\ref{fig:fig1}, right panel). The same effect has been already noticed by T2014 when
discussing the (less affected) $^{12}\rm{CO}(3-2)$ line profile. These authors estimate that the cloud affects their
measurements in the range 4-7 km/s $v_\mathrm{LSR}$. Integrating the moment 0 image over the regions of the image
with emission above $3\sigma$ (this region is delimited by the white dashed line contour in the central panel of Fig.~\ref{fig:fig1})
results in an integrated flux of $\mathrm{F_{^{12}CO}}=2.7\pm0.4$ Jy km/s. As in the case of the continuum data, this value
is an underestimation of the true flux due to the cleaning process. A Gaussian fit to the visibilities yields a larger value of
$\mathrm{F_{^{12}CO}}=3.0\pm0.4$ Jy km/s. Because of the cloud this value still represents a lower limit.

To determine the line center, we used public VLT/X-shooter spectra (ID: 089.C-0143) to derive the systemic radial
velocity of Sz\,91 (i.e., the center of the \co line) as $v_\mathrm{LSR}\approx3.4\pm 0.2$ km/s, in good agreement
with the results from \citet{melo2003} and T2014.

\section{Constrains on the star and disk}

In this section, we use our observations in combination with simple modeling to place constrains on
some of the disk's properties such as orientation, geometry, and mm slope. To complete our picture
of the Sz\,91 star$+$disk system we also derive the basic parameters of the central star.

\subsection{Stellar parameters}

First estimations of the Sz\,91 stellar parameters were obtained by \citet{hughes1994}, who used a distance
of $140\pm20$ au, an extinction $A_\mathrm{v}$ of 2 and a spectral type of M0.5. These authors derived a
stellar mass ranging from $M_{\star}=0.49-0.69 M_{\sun}$, an effective temperature of $T_{eff} = 3723$ K
and an age ranging from 5-7 Myr. \citet{romero2012} classified Sz\,91 as a M1.5 star using high resolution
spectra, in general agreement with the previous result. Using the latter spectral type, the most recent distance
estimate of $\sim$200 pc, and an extinction of $A_\mathrm{v}$ = 2, we derive the stellar parameters using the
stellar evolutionary models of \citet{siess2000}. We obtain a stellar radius of $R_\star = 1.46 R_\sun$, a temperature
of $T_{eff} = 3720$ K, a mass of $M_\star = 0.47 M_\sun$, and an age below 1 Myr\footnote{\citet{siess2000} notes
that their age estimations for stars below 1 Myr tend to be lower when compared with estimations from other models.}.
In what follows we use these values.

\subsection{$\alpha_{mm}$ slope}
\begin{figure}[t]
\center
\includegraphics[width = 1.0\linewidth,trim = 20 20 30 10]{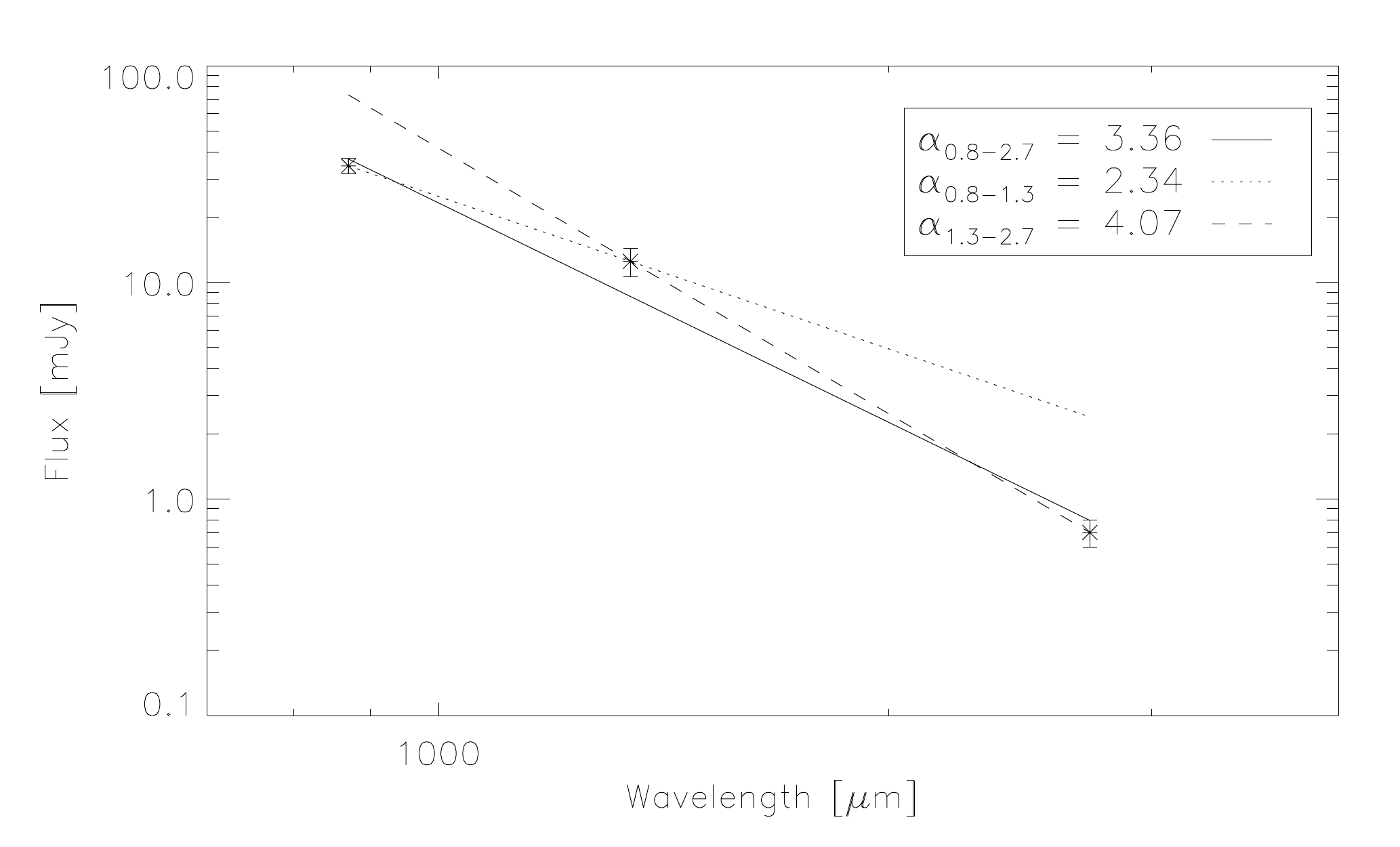}
\caption{Different fits to the mm-slope ($\alpha_{mm}$). It is not possible to simultaneously match
the three measurements (see text).}
\label{fig:fig3}
\end{figure}

At mm wavelengths, we can use the Rayleigh-Jeans approximation to express the flux as
$F_{\nu} \propto \nu^{\alpha_\mathrm{mm}}$. Assuming optically thin emission, the mm-slope is a function
of the dust opacity index $\beta$, i.e. $\alpha_\mathrm{mm} = 2 + \beta$. In particular, $\beta\sim2$ for the
interstellar medium (ISM) \citep[see][and references therein]{williams2011}. Trying to fit in this way the fluxes
at 0.85, 1.3 and 2.7 mm it is not possible to match the observed fluxes, as shown in Fig.~\ref{fig:fig3}. The
slope derived from a fit to the three fluxes results in $\alpha_\mathrm{0.8-2.7mm} = 3.36\pm0.14$ (solid line). 
Using the two shorter wavelengths we obtain $\alpha_\mathrm{0.8-1.3mm} = 2.34\pm0.40$ (dotted line), whereas
using the fluxes at 1.3 and 2.7 mm yields to $\alpha_\mathrm{1.3-2.7mm} = 4.07\pm0.29$ (dashed line).
This large $\alpha_\mathrm{1.3-2.7mm}$ would imply moderate or very low grain growth. On the other hand,
our derived $\alpha_\mathrm{0.8-1.3mm}$ implies $\beta_\mathrm{0.8-1.3mm} = 0.34\pm0.40$, which suggests
the opposite. This latter value agrees with the result by T2014 ($\beta = 0.5\pm0.1$) and with the average values
for a set of transition disks and protoplanetary disks derived by \citet{pinilla2014}. In addition, we conclude from
our models (see Sect. 5) that grains of at least 1 mm size are needed to match the SED at mm-wavelengths, implying
that grain growth is happening in Sz\,91 (see Sect. 5). One possible explanation of the different slopes is an
optical depth effect. Enhancements in the optical thickness could reduce the slope explaining the reduction in
$\alpha_{0.8-1.2mm}$ compared to the $\alpha_{1.3-2.7mm}$.
However, using a sample of 50 disks, \citet{ricci2012} find that low values of $\alpha_{mm}$ can be explained by
emission from optically thick regions only in the most massive disks if their sample. Given the relatively
low mass of Sz\,91 (Sect.~5), the good agreement with the results by T2014,  and the significant difference
between $\alpha_{0.8-1.3mm}$ and $\alpha_{1.3-2.7mm}$, we consider that an underestimation of the error bars
represents a more likely explanation of the broken $\alpha_{mm}$ slope and that $\alpha_\mathrm{0.8-1.3mm} = 2.34\pm0.40$
probably represents a realistic estimate of the true mm slope.

\subsection{Disk orientation}

Given the strong absorption observed towards the red-shifted channels, deriving the disk's center
and geometry by means of a Keplerian fit to the line profile as in e.g. \citet{mathews2012} would
result in large uncertainties. Instead, we determine the disk inclination ($i$), position angle (PA), and 
center by fitting different disk profiles to the continuum visibilities using the \textit{uvfit}/{\sc{MIRIAD}} task.

A ring morphology matches well our continuum observations and the results of T2014. Unfortunately,
\textit{uvfit} doesn't allow to adjust the thickness of the fitted ring. To test for the effect of using different disk
profiles, we performed three fits using a ring, a disk, and a Gaussian profile. Combining these results we
derived a PA of $18.2\degr\pm3.5$ (east of north) and an inclination ($i$) of $49.5\degr\pm3.5$. These results are
in agreement with those derived by T2014 ($\mathrm{PA}=17.5\degr\pm17.7\degr$, $i=40\degr\pm15\degr$). 
We find the disk's center to have an offset of $\Delta\alpha=-0\farcs29\pm0\farcs01$ and $\Delta\delta=-0\farcs06\pm0\farcs02$ 
with respect to the center of the ALMA field. This offset is consistent with the measured proper motion of (-20.3,-18.4)
mas $\mathrm{yr}^{-1}$ \citep[UCAC3, ][]{zacharias2010} when including the $\sim$0\farcs1 astrometric uncertainty
from our ALMA Cycle 0 observations and the 0\farcs06 uncertainty from our input coordinates \citep[2MASS catalogue, ][]{cutri2003}.
We corrected for this offset in our analysis of the disk.

\subsection{Outer disk}
\begin{figure}[t]
\center
\includegraphics[width = 1.0\linewidth,trim = 30 20 15 30]{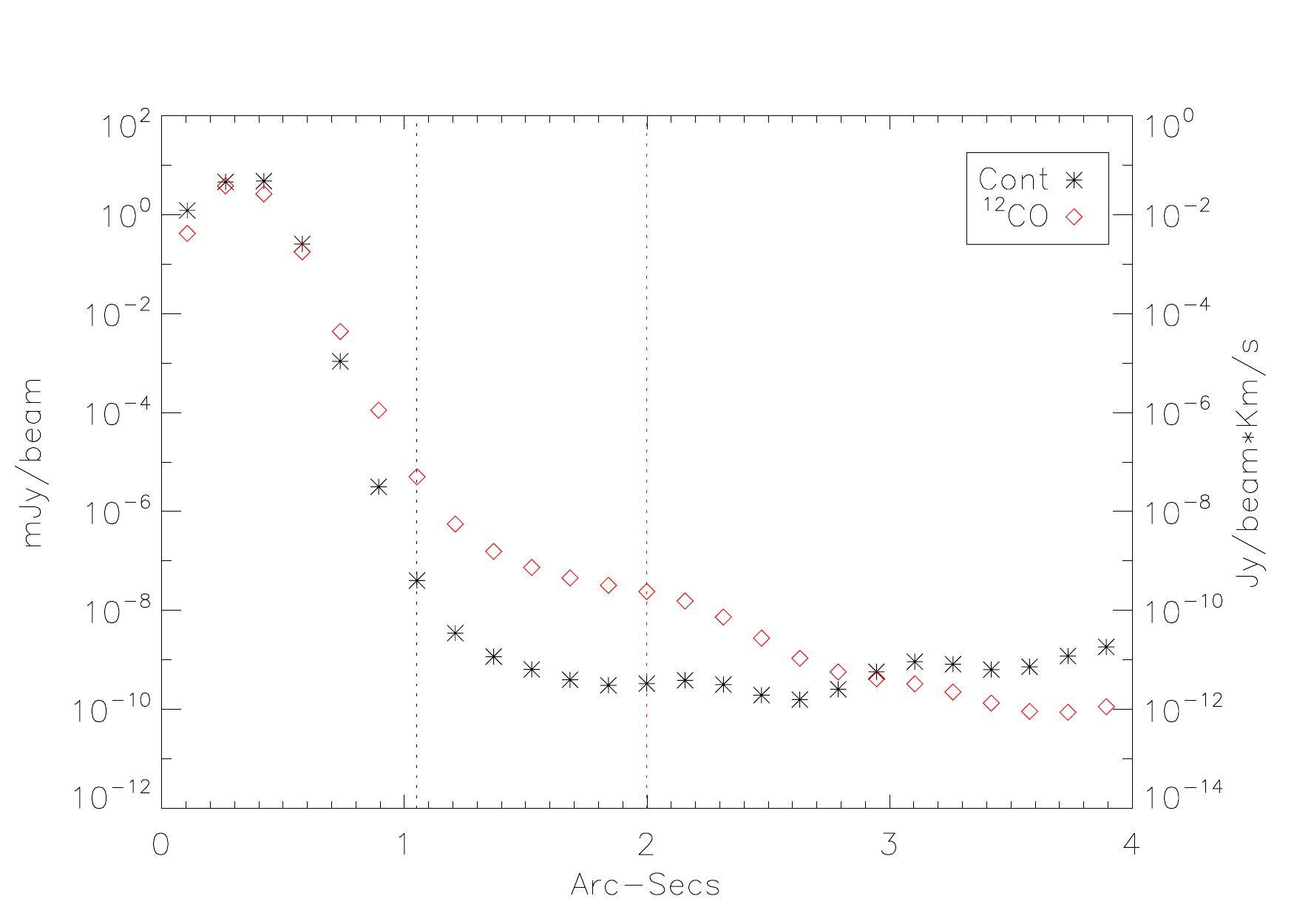}
\caption{Cut along the blue-shifted (i.e., not affected by cloud extinction) semi-major axis of the disk. The points are correlated (i.e., their
separation is smaller than the beam size along the disk's major axis). The vertical black and red dotted lines show the position at which
the emission is higher than $3\times$ the rms of the continuum and moment-0 \co images, respectively. The negative values are likely
artifacts from the cleaning algorithm.}
\label{fig:fig4}
\end{figure}

A quick inspection to the continuum (Fig.~\ref{fig:fig2}, left panel) and the \co moment-0 and moment-1 (Fig.~\ref{fig:fig2}, central panel)
images shows that the gaseous disk extends further out than the continuum disk. The resolution of our images along the major axis of
the disk is set by the beam size in that direction i.e. $0\farcs61$. Keeping this in mind, we produced a 3-pixel wide cut along the blue-shifted
(i.e., less affected by cloud emission) semi-major axis of the disk for the continuum and \co moment-0 images (Fig.~\ref{fig:fig4}). We
compute the fluxes every $0\farcs16$, which means that they are correlated (i.e., their separation is smaller than the beam size). Still,
we can use them to obtain information about the extension of the gaseous and dusty disks. In the plot, the vertical dotted lines indicate
the position at which the emission becomes significant (i.e., above $3\sigma$). The continuum emission becomes undetectable beyond
$\sim1\farcs1$ (i.e., $\sim$220 au) from the center, whereas the \co emission is still detected at $\sim2"$ ($\sim$400 au), i.e., almost 1.5
beams further out than the dusty disk. We therefore conclude that, although we are limited by the moderate resolution of the beam size,
the gaseous disk is undoubtedly detected at larger distance from the star than the dusty disk.  The outer radius of the dusty and gaseous
disk estimated by T2014 at $345$ GHz ($170\pm20$ au for the continuum, 420 au for the gas) matches notably well with our results. Our
limited spatial resolution doesn't allow us to probe for the sharp outer edge in the continuum disk observed in other disks, e.g., TW Hya
\citep{andrews2011} and HD\,163296 \citep{gregorio2013} that could be a consequence of radial drift \citep{birnstiel2013b}. 

\subsection{Inner disk}

\begin{figure*}[t]
\center
\includegraphics[width = 1.0\linewidth,trim = 5 10 30 120]{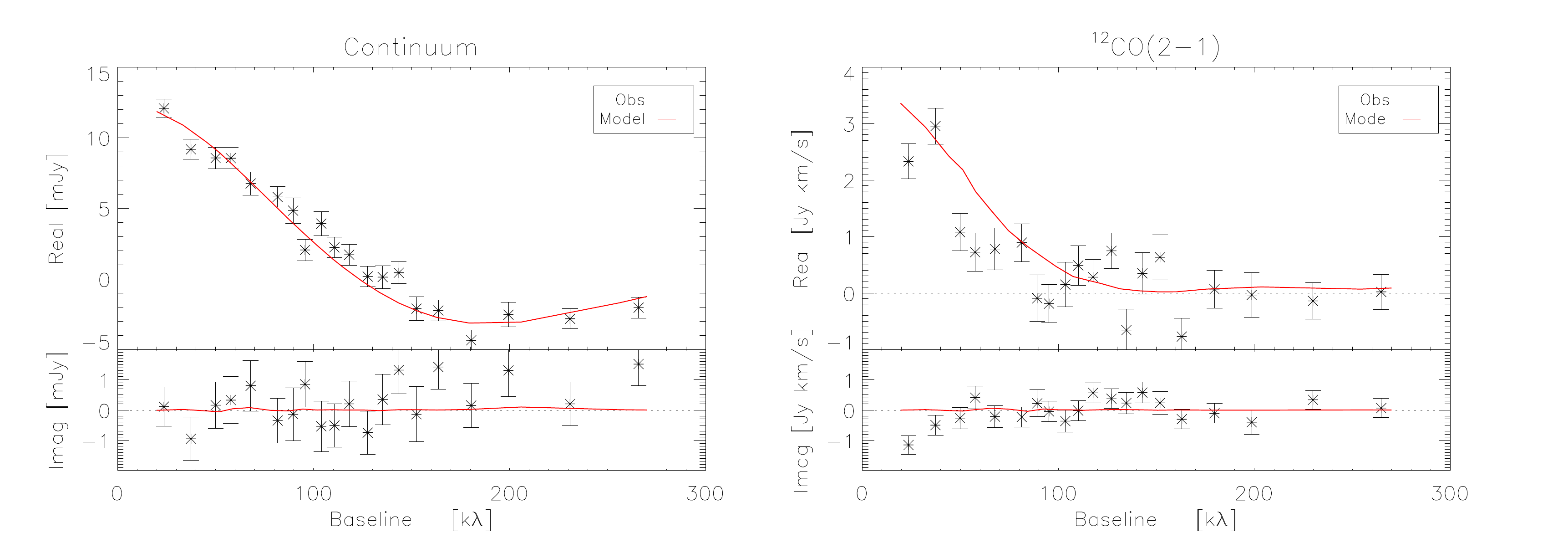}
\caption{Deprojected, radially binned real and imaginary visibilities as a function of $uv$ distance for the continuum (left) and the gas (right).
For the continuum, each bin shows the sum of the visibilities averaged over the continuum channels. For the \co, each bin shows the sum of
the visibilities over the channels containing significant emission (see text). The null in the continuum visibilities at $\sim125 \mathrm{k}\lambda$
indicates the presence of a large inner cavity in the mm-sized dust distribution. The red solid lines show the model discussed in Sect. 5.}
\label{fig:fig5}
\end{figure*}

The shape of the continuum cleaned image is indicative of the presence of an inner cavity. To explore the
partially resolved inner disk we here focus on the radially averaged deprojected ALMA visibilities.
We split the continuum and \co visibilities into 30 radial bins, ensuring that  all bins contain the same amount
of visibilities. This results in minimum and maximum bin sizes of 3.7 k$\lambda$ and 22 k$\lambda$, respectively, 
and represents a good compromise between $uv$ sampling and error bars. The real part of the binned visibilities
of our band-6 observations as a function of $uv$ distance are shown in Fig.~\ref{fig:fig5}. The steeper slope of the
\co visibilities shows that the detected gaseous disk is indeed more extended than the dusty disk, as already noticed
from the cleaned images. The null in the continuum visibilities at $\sim150 \rm{k}\lambda$ indicates a discontinuity
caused by the lack of mm-sized particles in the inner disk regions, or, in other words, the presence of a cavity
\citep[see ][]{hughes2007, andrews2011, hughes2009}. With our observations we cannot estimate the sharpness of
the inner wall \citep[see e.g., ][]{mulders2013}. In the next section we derive the cavity size in the continuum
by fitting the $uv$ plane.

The \co visibilities have lower S/N than the continuum and they are severely affected by the cloud.
This complicates the description of the gaseous disk in terms of the analysis of the visibility profile. However, if the
gaseous disk velocities are governed by Keplerian rotation, we can use the \co line profile to estimate how close to
the central star we detect gas using the projected observed velocity $v(r)_{Kep} = \sqrt{G M_{\star}/r}  \sin{i}$. Assuming
that the line center is at $v_\mathrm{LSR} = 3.4$ km/s, we detect \co emission down to $v_\mathrm{LSR} = 0.5$ km/s
with a S/N of $\sim$6 (Fig.~\ref{fig:fig2}). This corresponds to $-2.9$ km/s in the star's reference system.
Using this velocity, the stellar mass and inclination previously derived, and a distance of $200$\,pc, we conclude that our
observations probe the \co gas down to $\sim28$ au from the central star.

\section{Radiative transfer model}

Based on the obtained observational constraints, in what follows we develop an axisymmetric model for 
the disk around Sz\,91 using the radiative transfer code MCFOST \citep{pinte2006,pinte2009}. MCFOST
computes the dust temperature structure under the assumption of radiative equilibrium between the dust
and the local radiation field. Provided the proper set of parameters, MCFOST creates synthetic SEDs as
well as continuum and \co images. From these images we create synthetic interferometric observations
with the same $uv$ coverage as our observations, which are compared with our observations.

We start with a description of our model assumptions about the dust structure, composition, and
distribution. Then we use a genetic algorithm \citep{charbonneau1995} to identify a best-fit model that
can reproduce our observations.

\subsection{Dusty Disk Structure}

To account for the SED shape at mid-infrared wavelengths we define two  inner disk-regions: the innermost region
extending from the dust sublimation radius to $R_{out,1}$ which is severely dust-depleted (``Region 1") and accounts for
the fluxes at 12$\micron$ and below, and the intermediate region extending from $R_{in,2}$ to $R_{out,2}$ (``Region 2")
where some dust is emitting at 22 and 24 \micron.  There is no observational constrain on the surface density distribution
of these two regions, and therefore we adopt a simple power-law of the form
\begin{equation}
\Sigma (r) = \Sigma_{100} \left ( \frac{r}{100 \mathrm{ au}} \right)^{-p_s}
\label{eq:eq0}
\end{equation}

where $\Sigma_{100}$ is the surface density distribution at 100 au and $p_s$ is the power-law index.
We define the outer disk, ``Region 3", as the part of the disk extending from the cavity radius ($R_{\rm{cav}}$) 
outwards. For a number of disks where the gaseous disk is detected beyond the dusty disk as in Sz\,91 a
tapered-edge profile can naturally explain the observations \citep[e.g., ][]{hughes2008, williams2011, andrews2011,
mathews2012, cieza2012, zhang2014}. There are three exceptions to this behavior. To explain the high resolution
observations of  IM Lup, TW Hya and V4046 Sgr strong variations of the gas to dust ratio at large radii are required
\citep{panic2009,andrews2012,rosenfeld2013}. Our data do not have the high spatial resolution and sensitivity needed
to test for radial changes in the gas to dust ratio and therefore we adopt the tapered-edge prescription to describe the
outer disk. The surface density distribution in Region 3 is then defined as
\begin{equation}
\Sigma (r) = \Sigma_C \, r^{-\gamma} \, \mathrm{exp} \left [ - \left (\frac{r}{R_{\rm{C}}} \right) ^{2-\gamma} \right ],
\label{eq:eq1}
\end{equation}
where $R_{C}$ is the characteristic radius which defines where the tapered-edge begins to dominate over 
the power-law component,  $\Sigma_C$ is the surface density at $R_{C}$, and $\gamma$ corresponds
to the viscosity power-law index  in accretion disk theory \citep[$\nu\propto R^{\gamma}$,][]{lynden1974,
hartmann1998}. For each region we define the scale height $H(r)$ assuming that the dust follows a
Gaussian vertical density profile
\begin{equation}
H(r) = H_{0} (r/100\mathrm{au})^\psi,
\label{eq:eq2}
\end{equation}
where $\psi$ is the flaring parameter of the disk and $H_{0}$ is the scale height at $r=100$\,au.

\subsection{Dust composition and distribution}

We assume homogeneous spherical dust particles. The scattering opacities and phase functions, extinctions and
Mueller matrices are computed using Mie theory. For the composition we assumed compact astro-silicates \citep{mathis1989}
with a power-law size distribution \citep[$dn(a) \propto a^{-p}da$, with $p=3.5$ e.g., ][]{draine2006}. This assumption
requires minimum and maximum grain sizes of $a_{\mathrm{min}} =2\micron, a_{\mathrm{max}} =15\micron$ 
in region 1 to not overproduce the flux at 8 and 12 $\micron$ due to the silicate resonance bands. We note that using a
different grain composition as e.g. carbon grains it is also possible to match the SED using smaller values of $a_{\mathrm{min}}$
inside the cavity. In any case, the lack of emission above photospheric levels below $8\micron$ indicates that the innermost
region of the cavity must be severely depleted of dust particles. Region 2 is mainly constrained by the fluxes at 22, 24
and 70 $\micron$ and therefore many combinations of $a_{\mathrm{min}}, a_{\mathrm{max}}$, and the dust mass can reproduce
the observed fluxes. The minimum grain size, however, cannot exceed $a_{\mathrm{min}} =0.1\micron$ in order to match the
observed fluxes.

\subsection{Fitting procedure}

We use the stellar parameters derived in Sect. 4.1 and the inclination ($i$) and position angle (PA) derived in Sect. 4.3.
We focus on identifying a model that can describe the thermal structure of the disk (constrained by SED and
continuum observations) using the disk structure and dust parameters described in the previous sections. Regions 1
and 2 are poorly constrained and therefore we fix the power-law indexes of their surface density distribution to 
the standard value of $p_s = 1$ following \citet{andrews2007}. In Region 1 we fix the minimum and maximum
grain sizes to the values discussed in the previous section. Using this prescription we find that we need a \textit{maximum}
dust mass of $M_{dust,1} =  1\times10^{-4} \, M_\mathrm{\earth}$ to reproduce the SED shape below 12$\micron$.
We adopt this value to describe the dust content inside Region 1. A first exploration of the $H_{0}$ parameter in this
region reveals that it is completely unconstrained, and we adopt an arbitrary value of $H_{0_1} = 5$au. In Regions 1
and 2 we fix the minimum grain size to $a_{min} = 0.05\micron$, as in the interstellar medium. For Region 3 we 
explore the effects of varying the surface density parameter $\gamma$ within the range of $[-1,1]$. We find that
our data is equally well fitted with values ranging from $\gamma = [-0.2, 0.8]$ and use $\gamma = 0.3$. Similarly,
our data does not provide observational constraints on the flaring index ($\psi$) in any of the three regions. For all
three regions we adopt a typical value of $\psi=1.15$ from model results in , e.g. T Cha \citep{cieza2011}, IM Lup \citep{pinte2008} and HD\,163296
\citep{gregorio2013}.  Free parameters in our model approach are the outer radius of Region 1 and 2 ($R_{out_{1,2}}$),
the inner radius of Region 2 ($R_{in,2}$), and the cavity and characteristic radius of Region 3 ($R_{cav}, R_C$). We also
fit the dust masses $M_{dust_{2,3}}$, the maximum grain sizes $a_{max_{2,3}}$ and the scale heights $H_{0_{2,3}}$.

We explore the parameter space by means of a genetic algorithm. The procedure starts by randomly selecting a number
of models. The parameters of these models are defined as their ``genes''. Models producing a better match to our observational
dataset have the highest chance to  reproduce their parameter values (genes) into the next generation of models. The quality of
a fit is measured with the combined reduced $\chi^2$ given by the sum of reduced $\chi^2$ from SED and the continuum visibilities, i.e
$\chi^2 = \chi_{SED}^2 + \chi_{uv}^2$. We use a population of 50  ``parent" models and let them evolve through 50 generations,
resulting in a total of 2500 models. We find that after 35 generations the models quickly converge, and after 45 generations the variations
of each free parameter are within the 5\% of the average value of the parameter.

\subsection{Modeling results}

\begin{figure*}[t]
\center
\includegraphics[width = 1.0\linewidth,trim = 100 30 100 150]{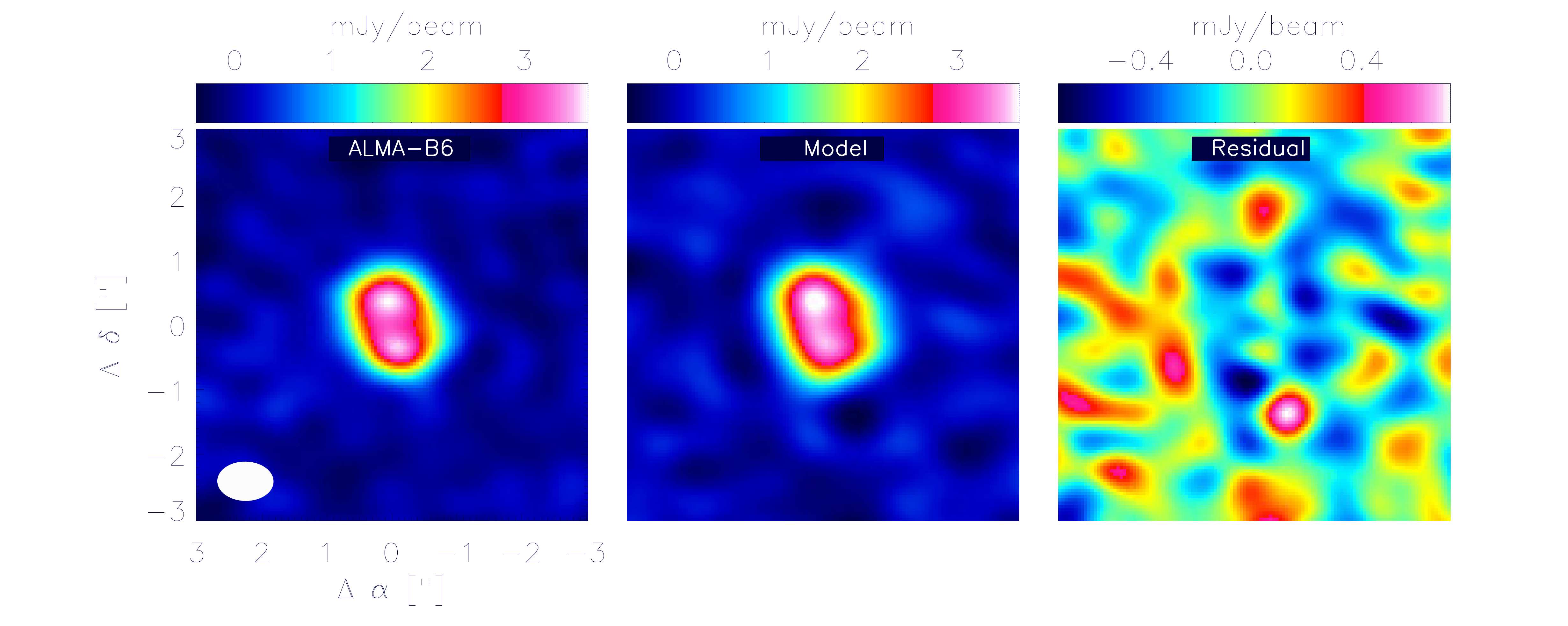}
\caption{Left: Band-6 continuum observations. Center: Synthetic, ALMA simulated images. Right: Residual image. Notice the different
color scale used in this figure.}
\label{fig:fig6}
\end{figure*}

\begin{figure}[t]
\center
\includegraphics[width = 1.0\linewidth,trim =50 40 40 50]{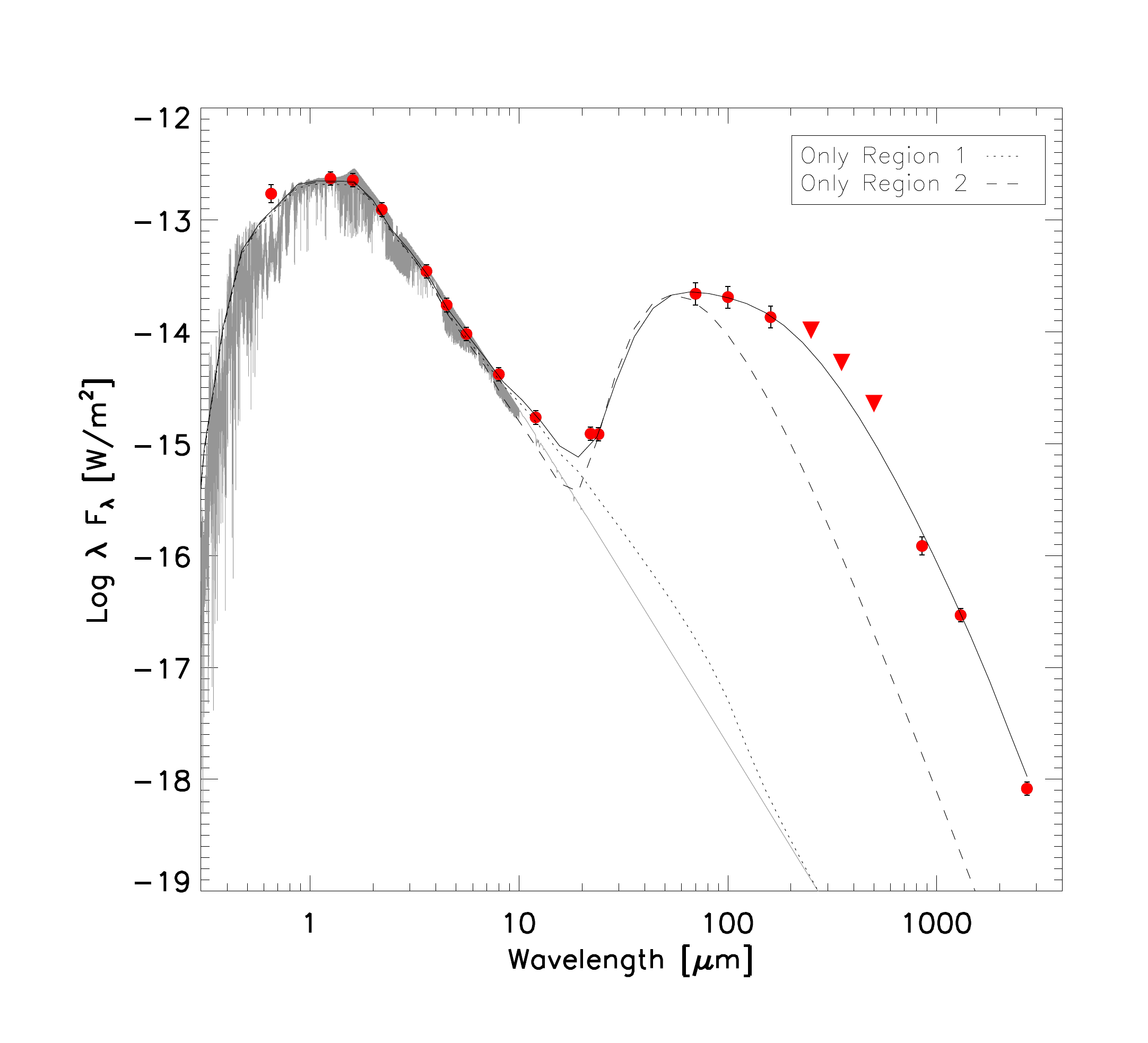}
\caption{Observed SED of Sz\,91, in red symbols. The simulated star's photosphere is shown in grey.
The model described in Sect. 5 is shown as a solid black line. The dotted and dashed lines 
represent the separate contributions of Region 1 and 2 of our model disk, respectively.}
\label{fig:fig7}
\end{figure}

\begin{figure}[t]
\center
\includegraphics[width = 1.0\linewidth,trim =50 20 20 20]{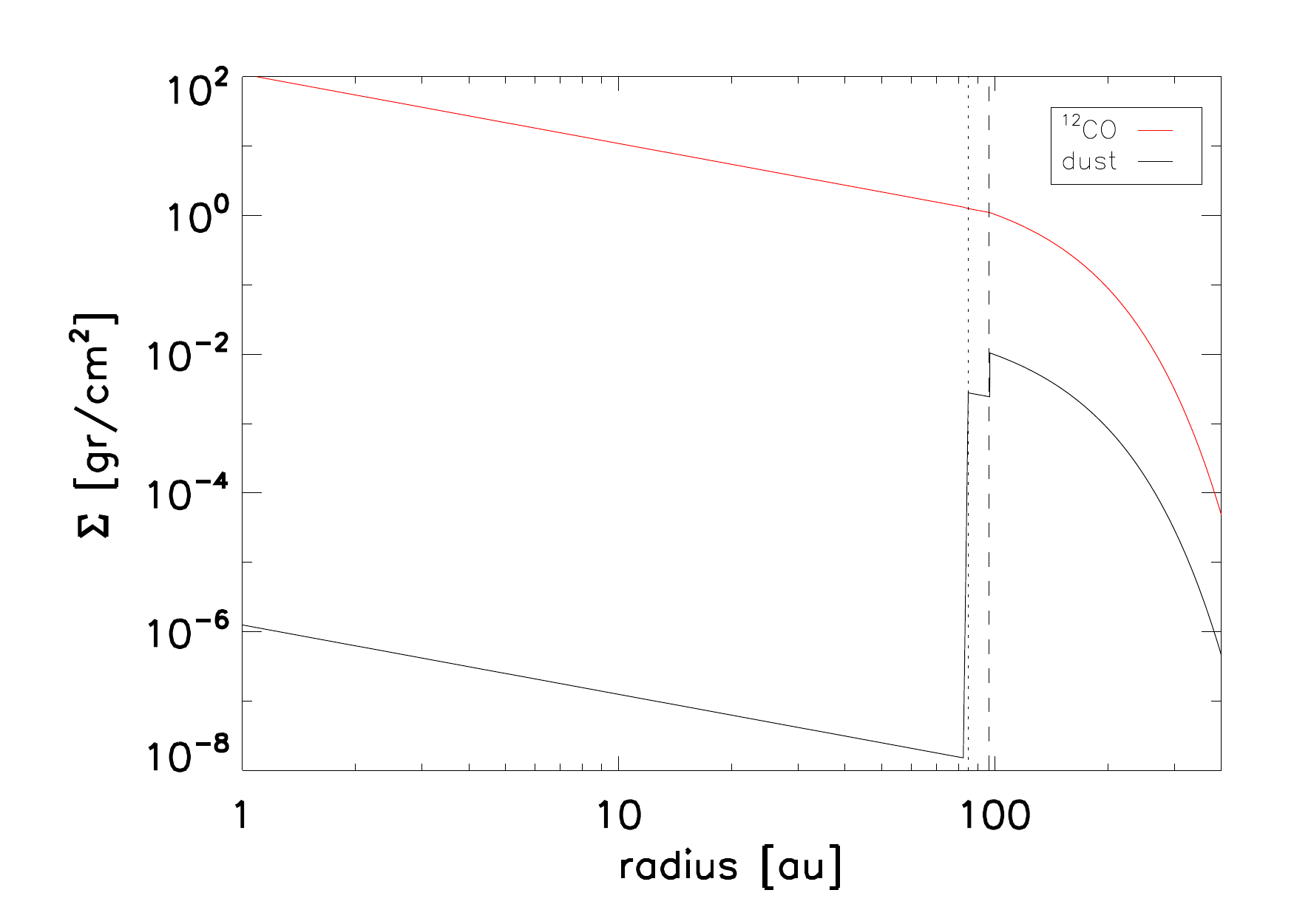}
\caption{Surface density distribution of the model. The dotted and dashed lines represent $R_{in,2}$ and $R_{cav}$, respectively.
The model extends down to 0.025 au (sublimation radius).}
\label{fig:fig8}
\end{figure}

We find a best-fit model that agrees well with our ALMA observations  and with the SED of Sz\,91.
Its parameters are summarized in Table~\ref{tab:tab3}.
\begin{table}[t]
\begin{center}
\caption{Model Results. Fixed parameters are listed above the line.}
\label{tab:tab3}
\begin{tabular}{c c}
\tableline\tableline
Parameter                                           &  Value                       \\ 
\tableline
	$T_{\star}$                                   & $3720$ K                    \\
	$R_{\star}$                                   & $1.46$ $R_{\sun}$      \\
	$M_{\star}$                                  & $0.47$ $M_{\sun}$      \\
	$d$                                               & 200 pc                         \\
	$i$                                                & $49.5\degr$                  \\
	$PA$                                            & $18.2\degr$                  \\
	$p$                                               & $3.5$                            \\
	$a_{\rm{min_1}},a_{\rm{max_1}}$ &$2, 15$\micron                                          \\
	$a_{\rm{min}_{2,3}}$                       &$0.05\micron$                                              \\
	$H_{0_1}$                                      &$5$ au                                                   \\
	$p_{s_{1,2}}$                                 & $1$                                                            \\
	$\gamma$                                      & $0.3$                                                         \\
	$\psi_{1,2,3}$                                 & $1.15$                                                       \\
	$M_\mathrm{dust_1}$                    &$1\times10^{-4}$ $M_\mathrm{\earth}$   \\
	$R_{\rm{1}}$                                   & $0.025$ au                                                \\

\tableline
	$a_{\rm{max_2}}$                                   &$5\micron$                                           \\
	$a_{\rm{max_3}}$                                   &$1000$\micron                                    \\
	$R_{out_{1}}$                                         & $85$ au                                                    \\
	$R_{in_{2}}$                                           & $85$ au                                                    \\
	$R_{out_{2}}$                                         & $97$ au                                                    \\
	$R_{cav}$                                               & $97$ au                                                    \\
	$R_{C}$                                                  & $100$ au                                                    \\
	$M_\mathrm{dust_2}$                            &$0.7$ $M_\mathrm{\earth}$                         \\
	$M_\mathrm{dust_3}$                            &$14.3$ $M_\mathrm{\earth}$                        \\
	$H_{0_2}$                                               &$5$ au                                                  \\
	$H_{0_3}$                                               &$10$ au                                                \\
	$\mathrm{log}(\Sigma_{100_1})$           &$-8.20$ [$gr/cm^2$]\tablenotemark{a}             \\
	$\mathrm{log}(\Sigma_{100_2})$           &$-2.95$ [$gr/cm^2$]\tablenotemark{a}             \\
	$\mathrm{log}(\Sigma_{C})$                  &$-1.98$ [$gr/cm^2$]\tablenotemark{a}             \\

\tableline
\end{tabular}
\tablenotetext{a}{Derived values.}
\end{center}
\end{table}
The most significant result of our modeling is that a large cavity of $\sim97$ au in the dust grain distribution
is required to mach our observations. This is clearly illustrated by the null in the real part of the continuum deprojected
visibilities as shown in Fig.~\ref{fig:fig5} (left panel). For direct comparison
with our ALMA band-6 images we created synthetic ALMA observations using the {\sc simalma/casa}\footnote{More
information about the CASA package can be found at: http://casa.nrao.edu/} package. We created an antenna configuration file that 
mimics our observations using the {\sc buildConfigurationFile/Analysis Utilities}\footnote{Analysis Utilities is an external package provided
by NRAO to complement the CASA utilities. It can be found at: http://casaguides.nrao.edu/index.php?title=Analysis\_Utilities} task. 
We then run the simulation using the same PWV, exposure time and hour angle as in our observations. The synthetic ALMA image
of our model for Sz\,91 and the residuals obtained by subtracting the synthetic image from the real observations are shown in 
Fig.~\ref{fig:fig6} (central and right panels). The residual image has an rms of 0.1 mJy/beam similar to the rms of the observations.
Interestingly, our simulated image also shows a brightness difference between the northern and southern lobes of the disk, with the
northern lobe being the brighter one (although our MCFOST model image is axisymmetric). To investigate this feature we repeated 
the SIMALMA/CASA simulations using different $uv$ coverages and PWV. We find that the asymmetry is a consequence of the low
S/N of the visibilities at longer baselines as increasing the $uv$ coverage at larger baselines or decreasing the PWV of the simulated
ALMA observations removes the apparent difference in brightness.
\\
The SED and surface density distribution of the model are shown in Figs.~\ref{fig:fig7} and \ref{fig:fig8}, respectively.
In Fig.~\ref{fig:fig7} we additionally show the individual contributions of Region 1 and Region 2.
\subsection{Model uncertainties}
\begin{figure*}[t]
\center
\includegraphics[width = 1.0\linewidth,trim = 80 0 95 50]{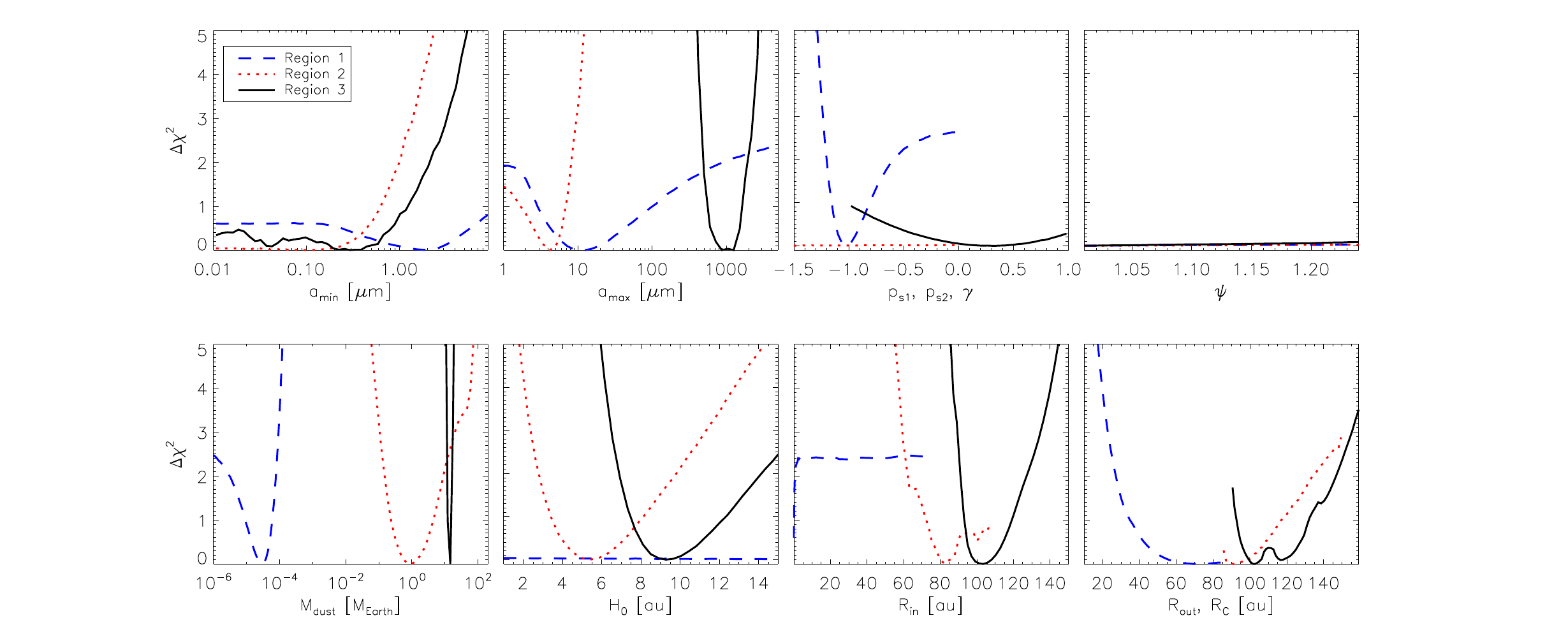}
\caption{$\Delta\chi^2$ (i.e., the difference between the reduced $\chi^2$ and its minimum) of the fixed parameters in our
modeling (see text). Blue dashed line corresponds to Region 1, red-dotted and black-solid lines indicate Regions 2 and 3,
respectively.}
\label{fig:fig9}
\end{figure*}

To estimate which model parameters are well constrained from our observational dataset we proceed as follow.
We fix all but one parameter and run a grid of models exploring the \textit{local} parameter space. We then derive the
reduced $\chi^2$ of the models forming this local grid. This procedure is repeated for all free parameters.
Keeping in mind that this is equivalent to assume that all parameters are uncorrelated (which is clearly not the case), 
this approach allow us to get a first idea of which parameters are clearly constrained by our observations without spending
unreasonable amounts of computing time. 
The results of this exploration are shown in Fig.~\ref{fig:fig9}. These plots illustrate the severe degeneracy of 
most of the parameters describing Regions 1 and 2 in our model disk. In contrast, we find that the parameters such
as $M_{dust}$, $a_{max}$ and the cavity size ($R_{cav}$) in Region 3 are more constrained by our observations. 
In particular, values of $R_{cav}$ below 90 au are very difficult to reconcile with the ALMA observations. According
to our model, most of the disk's mass (10 $M_{\earth}, \sim70\%$ of $M_{dust,3}$) is concentrated in a ring ranging
from 97 to 140 au.

\subsection{Gaseous Disk}

Our observations trace the optically thick \co line which is severely affected by the cloud. We therefore only present a
simple model of the gas component of the disk around Sz\,91 that is able to match the \co line profile reasonably well using
standard assumptions. We only consider the blue-shifted (less absorbed) side of the line. Given the sparse observational
constrains on the gaseous disk, we build this model upon the best-fit model for the dusty disk obtained in the previous section.
The lack of measurements of the isotopologues $^{13}\mathrm{CO}$ and $\mathrm{C^{18}O}$ prevents us
from estimating the gas to dust ratio and total gas mass as in \citet{williams2014}. We fix the gas to dust ratio in the outer region (Region 3)
to the standard value of 100, and we ensure that the \co freezes-out
at 20 K \citep{qi2004, gregorio2013}. We assume that the gas and the dust have the same temperature (local thermal equilibrium,
LTE). For the \co abundance we use the standard $\mathrm{H}_{2}$ / \co ratio ($10^{4}$), which is a reasonable value for
protoplanetary disks as confirmed by \citet{france2014}. The gas is assumed to be in Keplerian rotation with an inner radius of at
least 28 au (in agreement with our observations). Assuming that the gas is in hydrostatic equilibrium in this direction, the
gas profile is described by
\begin{equation}
\rho(r,z) = \rho(r,0) \mathrm{exp} \left [ - \frac{z^2}{2 {H(r)}^{2}} \right ].
\label{eq:eq2}
\end{equation}
The scale height of the gas component $H(r)$ is given by 
\begin{equation}
H(r) = \sqrt{\frac{r^3 k_{\mathrm{B}} T(r)}{G M_{\star} \mu m_{H}}},
\label{eq:eq2}
\end{equation}
where $k_{\mathrm{B}}$ is the Boltzmann constant, $G$ is the gravitational constant and $m_{H}$ is the Hydrogen
mass. We also take into account the thermal broadening of the velocities (defined as
$V_{th} = \sqrt{2 k_\mathrm{b} T_\mathrm{CO} / m_\mathrm{CO}}$, where $m_\mathrm{CO}$ is the molecular mass
of the \co) and the effect of turbulence ($v_\mathrm{turb}$) on the \co emission along the line of sight. 

\begin{figure}[t]
\center
\includegraphics[width = 1.0\linewidth,trim = 25 10 20 0]{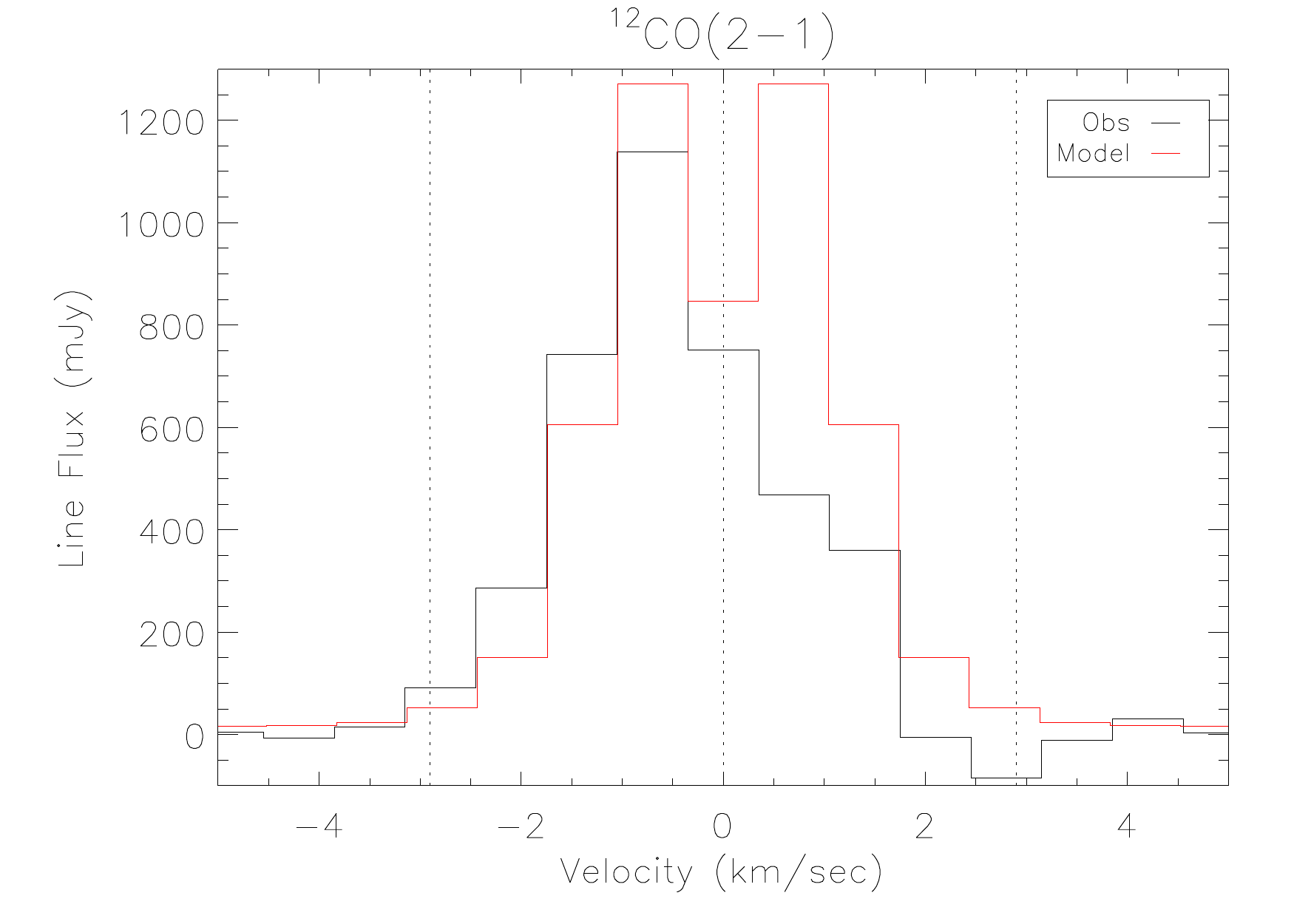}
\caption{Observed \co line profile, in black histogram. The velocity scale is centered at the reference system of the star,
using $v_\mathrm{LSR}=3.4$ km/s. The profile was computed by integrating the flux contained in the $>3\sigma$
region of the moment-0 image (white dashed contour in central panel of Fig.~\ref{fig:fig2}) in each channel. The vertical
dotted lines show the center of the line and the highest velocity bin observed at $>3\sigma$ significance, corresponding
to -2.9 km/s. The vertical dotted line at 2.9 km/s highlights the strong absorption-like effect on the red-shifted side of the \co line.
The red histogram shows the gas model presented in Sect.~5.}
\label{fig:fig10}
\end{figure}

Under these assumptions, we manually explore different models by varying the turbulence velocity ($v_\mathrm{turb}$)
within a range of [10,300] m/s \citep[in accordance with the observations discussed by][]{hughes2011} and the gas mass inside
the dust depleted regions (Region 1 and 2).
We find a model that agrees relatively well with our observations using a continuous gaseous disk (i.e. without depletion
or inner cavity) with $v_\mathrm{turb} = 120$ m/s. The radially binned deprojected visibilities and the surface density of
this model are shown in Fig.~\ref{fig:fig5} (right panel) and Fig.~\ref{fig:fig8} (red line), respectively. The integrated \co
line profile of the model, along with the real observations, is shown in Fig.~\ref{fig:fig10}.

\section{Discussion}
\subsection{Are three zones required?}

Assuming three different disk regions we find a model that reproduces reasonably well 
our observational dataset. As a separate exercise we also tried to fit a disk model composed by 
only two regions (the inner cavity and the outer disk). After running the genetic search we could
not find a good match to our observations with such a disk structure. This result is in agreement
with the findings of T2014, who need an extra component, either a circumplanetary disk or a narrow,
optically thick inner ring, to reproduce the fluxes at 22 and 24$\micron.$ 

\subsection{Model limitations}

The local parameter exploration (Fig.~\ref{fig:fig9}) represents a useful approach to estimate how
sensitive are the model parameters to our observations. We find that most of the parameters describing
Region 1 and 2 remain unconstrained from our observations. For instance, in our representative model
there are no empty gaps as we find $R_{out,1} = R_{in,2}$ and $R_{out,2} = R_{cav}$, although other
values for these parameters (i.e., an empty gap between Region 1 and 2, or a dust empty innermost region)
could  equally reproduce our observations (as indicated by the $\Delta \chi^2$ profile in the $R_{in}$ and
$R_{out}$ panels in Fig.\ref{fig:fig9}). For all 3 regions we find that the the exponents of the surface density
distributions are completely unconstrained by our observations. The same happens with the flaring index $\psi$,
as expected if the emission is optically thin at these wavelengths. 
However, the exercise we performed does not allow to derive proper statistical
errors of the fitted parameters as we did not investigate correlations between the parameters. The latter would require
to map the entire multidimensional parameter space which is beyond the scope of this paper and which would require to spend
an unreasonable amount of computing time. We reserve this exercise for the future, when data with more sensitivity and much
better angular resolution will be available. 
Regarding the gas model, we find that it is difficult to reproduce the observed \co profile with our model. Overall
we see that compared to the observations the model shows a lack of emission at higher velocities, and an
excess at lower velocities. Even using a non-depleted gaseous disk our model produces less flux at higher velocities
than the observations. We note that this does not exclude the possibility of a gas-depleted cavity 
\citep[as found for HD\,142527,][]{perez2014},
since we also find that it is possible to produce a similar \co profile modeling a gas-depleted cavity using higher values of
turbulence, or using different surface density profiles for the gaseous and dusty disks (i.e., different $\gamma$ for the gas
and dust disks).

\subsection{Cavity-making mechanism}

With a size of 97 au in radius, Sz\,91 has by far the largest the inner cavity observed in a transition disk around a T Tauri star.
This cavity size and accretion rate ($7.9\times10^{-11}\mathrm{M_{\sun}}\,\mathrm{yr}^{-1}$)
exclude photoevaporation alone as the mechanism responsible for the cavity 
\citep[e.g., ][]{rosotti2013, owen2012}. Similarly, studies of grain growth conclude that grain growth alone cannot produce the large
cavities resolved by interferometric mm observations \citep{birnstiel2013a}. \citet{romero2012} excluded a stellar companion down
to separations of $\sim30$\,au with a contrast of $\Delta K \sim3.3$. Using NextGen models \citep[e.g.][]{allard1997}, an age
of 1 Myr and the stellar mass derived in Sect. 4.1 translates into a sensitivity close to the brown dwarf limit of $\sim80~\mathrm{M_{J}}$.
In addition, \citet{melo2003} found no evidence for a close-in binary companion around Sz\,91 in their 3-years radial velocity survey
(down to masses of $\approx0.2 \mathrm{M}_{\odot}$). However, we cannot yet exclude that a low mass stellar companion is carving
the cavity in the disk around Sz\,91.
The other possible mechanism is planet formation. In fact, our direct detection of gas inside of the large dust-depleted cavity, the presence
of $\micron$-sized dust inside the cavity, and the large cavity size indicated by the continuum visibilities matches well several predictions
of models for \textit{multiple} (Jupiter-sized) planet formation \citep[e.g.][]{lubow2006, dodson2011,zhu2011}.

\subsection{Inner disk structure: planet formation signature?}
There are at least four transition objects with detectable (but different) accretion rates that can be explained by an inner disk structure
similar to the one derived for Sz\,91. DM\,Tau and RX\,J1615.3-3255 have resolved cavities and require an empty disk very close to the
star because of the lack of NIR excess, but some dust in the outer regions of the cavity to account for the MIR emission \citep{andrews2011}.
However, a direct comparison with Sz\,91 is difficult because the modeling approach adopted by \citet{andrews2011} is different from ours.
RX\,J1633.9-2442 and [PZ99] J160421.7-213028 have been analyzed using similar tools than those used in this work for Sz\,91. In both
systems, an empty inner disk and some dust inside the main cavity \citep{cieza2012,mathews2012,zhang2014} are needed to explain 
the mm-observations and SED. Interestingly, the mm-sized dust distribution in [PZ99] J160421.7-213028 is concentrated in a ring similar
to what we find for Sz\,91 \citep{zhang2014, mathews2012}. \citet{pinilla2012} concludes that this ring-shaped dusty structure is a product
of the planet-disk interaction. 

While in all five cases strong parameter degeneracies of the models do not allow us to constrain the exact configuration of the dust inside
the cavity, photoevaporation and grain growth can be ruled out as the mechanisms carving the inner hole of these disks. Furthermore, for
DM\,Tau, RX\,J1633.9-2442, and [PZ99] close binarity can also be excluded and planet-formation has been suggested to explain both the
cavities and the presence of dust inside the cavity \citep[e.g., ][]{rice2006,cieza2012,mathews2012}. As suggested by \citet{cieza2012},
these inner structures are likely simple approximations to the complex structures predicted by hydrodynamical models of planet formation
\citep{dodson2011}, which have now been directly observed \citep[e.g.,][]{casassus2013, quanz2013} in a few systems.

\section{Conclusions}

We presented ALMA band-6 and band-3 observations of Sz\,91 which clearly show that there must be a very large inner
cavity in the disk around Sz\,91. We also detect \co gas inside the cavity down to at least 28 au. We detect gas at larger radii
(400 au) than the dust (220 au). 
We used a genetic algorithm to construct an MCFOST model of the disk around Sz\,91 explaining our ALMA observations and
the SED of the system. We find that a three-zone model is required to fully explain the observations. The inner region of the disk is 
significantly depleted of dust particles, and there are $\approx0.7\rm{M}_{\earth}$ of dust confined within the 85-97 au region. 
The exact spatial distribution of the dust in these two regions is not constrained by the available observations but our $uv$-fit
to the data indicates that the dust depleted region extends to $\sim97$\,au. This implies that there must be significant amounts of gas inside the dust
depleted inner regions. The bulk of the dust mass is located in a ring extending from 97 au to 140 au. The different outer radius
observed for the \co and the dust can be explained by an exponential tapered surface density profile.

A yet undetected, very low mass companion (below $0.2\mathrm{M_{\sun}}$) in the inner 30 au, could create
the large cavity observed in Sz\,91. However, the size of the disk's cavity and the presence of gas and small dust
particles inside the cavity agree remarkable well with the theoretical predictions for \textit{multiple}, Jupiter-sized,
planet formation. Considering its spectral type and low stellar mass,
Sz\,91 has an extremely large inner cavity when compared with other transition disks \citep[e.g.][]{andrews2007}.
Finally, we find that similar structures of the inner regions of dusty disks are also observed in at least four other
planet-forming candidates transition disks. It is tempting to interpret the inner disk structure and large cavity size 
as a signpost of forming planets inside these disks.

\acknowledgments
The authors thank the anonymous referee for providing very constructive comments, which significantly
helped to improve the paper. We are grateful to the ALMA staff.
This research was funded by the Millenium Science Initiative,
Chilean Ministry of Economy, Nucleus RC130007. H.C. and C.C. thank for support from 
ALMA/CONICYT (grants 31100025 and 31130027). 
CC also acknowledges support from CONICYT-FONDECYT grant 3140592. 
MRS acknowledges support from FONDECYT (grant 1141269).
C.P. acknowledges funding from the European Commission's 7$^\mathrm{th}$ Framework Program (contract PERG06-GA-2009-256513) and
Agence Nationale pour la Recherche (ANR) of France (contract ANR-2010-JCJC-0504-01). 
L.C. was supported by ALMA-CONICYT grant number 31120009 and CONICYT-FONDECYT grant number 1140109.
S.C. acknowledges support from FONDECYT grant 1130949. 
J.P.W. acknowledges support from NSF grant AST-1208911 and NASA grant NNX15AC92G.
P.R. acknowledges support from ALMA-CONICYT 31120006 and FONDECYT grant 3140634.
This publication makes use of
data products from the Two Micron All Sky Survey, which is a joint project
of the University of Massachusetts and the Infrared Processing and Analysis
Center/California Institute of Technology, funded by the National Aeronautics
and Space Administration and the National Science Foundation.



{\it Facilities:} \facility{ALMA}




\clearpage


\end{document}